\documentclass[useAMS,usenatbib]{mn2e}

\usepackage{graphicx}
\usepackage{scalefnt}
\usepackage{amssymb}
\usepackage{upgreek}

\newcommand{\ud}{\mathrm{d}}

\def \etal {et~al.~}
\def \chisq  {\ifmmode  \chi^2   \else  $\chi^2$  \fi}  
\def \spose#1{\hbox  to 0pt{#1\hss}}  
\def \lta{\mathrel{\spose{\lower 3pt\hbox{$\sim$}}\raise  2.0pt\hbox{$<$}}}
\def \gta{\mathrel{\spose{\lower  3pt\hbox{$\sim$}}\raise 2.0pt\hbox{$>$}}}

\def \kms {\ifmmode  \,\rm km\,s^{-1} \else $\,\rm km\,s^{-1}  $ \fi }
\def \kpc {\ifmmode  {\rm~kpc}  \else ${\rm~kpc}$\fi}  
\def \pc {\ifmmode  {\rm~pc}  \else ${\rm~pc}$ \fi  }  
\def \Gyr {\ifmmode  {\rm~Gyr}  \else ${\rm~Gyr}$\fi}
\def \Msun {\ifmmode M_{\odot} \else $M_{\odot}$ \fi} 
\def \Lsun {\ifmmode L_{\odot} \else $L_{\odot}$ \fi} 
\def \Rsun {\ifmmode R_{\odot} \else $R_{\odot}$ \fi} 
\def \Msunpyr {\ifmmode M_{\odot}{\rm~yr}^{-1} \else $M_{\odot}{\rm~yr}^{-1}$ \fi} 
\def \hMsun {\ifmmode h^{-1}\,\rm M_{\odot} \else $h^{-1}\,\rm M_{\odot}$ \fi}

\def \LCDM {\ifmmode \Lambda{\rm CDM} \else $\Lambda{\rm CDM}$ \fi}
\def \sig8 {\ifmmode \sigma_8 \else $\sigma_8$ \fi} 
\def \OmegaM {\ifmmode \Omega_{\rm M} \else $\Omega_{\rm M}$ \fi} 
\def \OmegaL {\ifmmode \Omega_{\rm \Lambda} \else $\Omega_{\rm \Lambda}$\fi} 
\def \Deltavir {\ifmmode \Delta_{\rm vir} \else $\Delta_{\rm vir}$ \fi}
\def \rhocrit {\ifmmode \rho_{\rm crit} \else $\rho_{\rm crit}$ \fi}
\def \rhou {\ifmmode \rho_{\rm u} \else $\rho_{\rm u}$ \fi}
\def \zc {\ifmmode z_{\rm c} \else $z_{\rm c}$ \fi}

\def \rhos {\ifmmode \rho_{\rm s} \else $\rho_{\rm s}$ \fi} 
\def \rs {\ifmmode r_{\rm s} \else $r_{\rm s}$ \fi} 
\def \cvir {\ifmmode c_{\rm vir} \else $c_{\rm vir}$ \fi} 
\def \Rvir {\ifmmode r_{\rm vir} \else $R_{\rm vir}$ \fi}
\def \Vvir {\ifmmode V_{\rm  vir} \else  $V_{\rm vir}$  \fi} 
\def \Mvir {\ifmmode M_{\rm  vir} \else $M_{\rm  vir}$ \fi}  
\def \Nvir {\ifmmode N_{\rm  vir} \else $N_{\rm  vir}$ \fi}  
\def \Jvir {\ifmmode J_{\rm vir} \else $J_{\rm vir}$ \fi} 
\def \Evir {\ifmmode E_{\rm vir} \else $E_{\rm vir}$ \fi} 
\def \vvir {\ifmmode v_{\rm vir} \else $v_{\rm vir}$ \fi} 
\def \lam {\ifmmode \lambda  \else $\lambda$ \fi} 
\def \lamp {\ifmmode \lambda^{\prime} \else $\lambda^{\prime}$  \fi} 
\def \Vmax {\ifmmode V_{\rm  max} \else  $V_{\rm max}$  \fi} 
\def \Mdm {\ifmmode M_{\rm  dm} \else $M_{\rm  dm}$ \fi}

\def \Mgas {\ifmmode M_{\rm gas} \else $M_{\rm gas}$ \fi} 
\def \Mcg {\ifmmode M_{\rm cg} \else $M_{\rm cg}$\fi} 
\def \Mhg {\ifmmode M_{\rm hg} \else $M_{\rm hg}$ \fi} 
\def \Mdisc {\ifmmode M_{\rm disc} \else $M_{\rm disc}$ \fi} 
\def \Md {\ifmmode M_{\rm d} \else $M_{\rm d}$ \fi} 
\def \Mda {\ifmmode M_{\rm d,0\%} \else $M_{\rm d,0\%}$ \fi} 
\def \Mdb {\ifmmode M_{\rm d,20\%} \else $M_{\rm d,20\%}$ \fi} 
\def \Mdc {\ifmmode M_{\rm d,40\%} \else $M_{\rm d,40\%}$ \fi} 
\def \md {\ifmmode m_{\rm d} \else $m_{\rm d}$ \fi} 
\def \Mb {\ifmmode M_{\rm b} \else $M_{\rm b}$ \fi} 
\def \Mbh {\ifmmode M_{\rm b,pri} \else $M_{\rm b,pri}$ \fi} 
\def \Mbs {\ifmmode M_{\rm b,sat} \else $M_{\rm b,sat}$ \fi} 
\def \zo {\ifmmode z_{0} \else $z_{0}$ \fi} 
\def \rd {\ifmmode r_{\rm d} \else $r_{\rm d}$ \fi}
\def \rg {\ifmmode r_{\rm g} \else $r_{\rm g}$ \fi}
\def \rb {\ifmmode r_{\rm b} \else $r_{\rm b}$\fi}
\def \rs {\ifmmode r_{\rm s} \else $r_{\rm s}$\fi}
\def \rc {\ifmmode r_{\rm c} \else $r_{\rm c}$\fi}
\def \rvir {\ifmmode r_{\rm vir} \else $r_{\rm vir}$\fi}
\def \rbh {\ifmmode r_{\rm b,pri} \else $r_{\rm b,pri}$ \fi} 
\def \rbs {\ifmmode r_{\rm b,sat} \else $r_{\rm b,sat}$ \fi}

\title[Tidal stripping and morphology]{The dependence of tidal stripping efficiency on the satellite and host galaxy morphology}
\author[J. Chang et al.]{Jiang Chang$^{1,2}$\thanks{jchang@mpia.de}, Andrea V. Macci\`o$^2$, Xi Kang$^1$\\
$^1$ Purple Mountain Observatory, the Partner Group of MPI f\"ur Astronomy, 2 West Beijing Road, Nanjing 210008, China\\
$^2$ Max-Planck-Institut f\"ur Astronomie, K\"onigstuhl 17, 69117 Heidelberg, Germany\\ 
 }

\begin{document}

\pagerange{\pageref{firstpage}--\pageref{lastpage}} \pubyear{---}

\maketitle

\label{firstpage}

\begin{abstract}

In  this paper  we study  the  tidal stripping  process for  satellite
galaxies orbiting around a massive  host galaxy, and focus on its dependence
on the morphology  of both satellite and  host galaxy.  For this
purpose, we use three  different morphologies for the satellites: pure
disc, pure bulge  and a mixture bulge+disc.
Two morphologies are  used for the  host galaxies: bulge+disc and pure bulge.
We find that while the spheroidal stellar component experiences a
constant power-law like mass removal, the disc is exposed to an exponential 
mass loss when the tidal radius of the satellite is of the
same order of the disc scale length. This dramatic mass loss
is able to completely remove the stellar component on time scale
of 100 Myears.
As a consequence two satellites with the same stellar and dark matter
masses, on the same orbit could either retain considerable fraction of their stellar mass after 10 Gyrs 
or being completely destroyed, depending on their initial stellar morphology.
We find that there are two characteristic time scales describing the beginning
and the end of the disc removal, whose values are related to the size of the
disc. This result can be easily incorporated in semi-analytical models. We also find
that the host morphology and the orbital parameters also have an effect on the
determining the mass removal, but they
are of secondary importance with respect to satellite morphology.
We conclude that satellite morphology has a very strong effect on the efficiency
of stellar stripping and should be taken into account in modeling galaxy
formation and evolution.

\end{abstract}


\section{Introduction}

The current model for structure  formation in the Universe is based on
the cold dark  matter (CDM) theory (White \&  Rees 1978; Blumenthal et
al. 1984). In this hierarchical  scenario small dark matter halos
form first, and then they subsequently merge to form larger ones.  At the
same time,  gas cools and collapses  into the potential  well of these
dark matter haloes  where star formation takes place,  giving rise to the
first   protogalaxies.    When  dark   matter   haloes  merge,   these
protogalaxies also  merge; depending on  the mass ratio  the satellite
galaxy  can either  rapidly merge  with  the central  object or  orbit
around it  for a  consistent amount of  time (e.g., Chandrasekhar 1943;
Binney \& Tremain 2008; Jiang et al. 2008; Boylan-Kolchin \etal 2008).

In the second case the  satellite galaxy will experience a progressive
mass loss, both  in its dark matter and  stellar component (e.g., Mayer
\etal  2001a;  Klimentowski  \etal   2007;  Penarrubia  \etal  2008;
Kazantzidis  \etal 2011).  Observations  have confirmed  this scenario
with the  discovery of streams and complex  stellar structures (shells
and  cusps) associated  with  the accretion  and  tidal disruption  of
satellites galaxies in our own  Galaxy (e.g., Ibata et al. 1994; Yanny
et al.   2000; Newberg et al.   2002; Majewski et al.  2003; Martin et
al. 2004; Martınez-Delgado et al. 2005; Belokurov et al. 2006), in the
Andromeda galaxy (Ibata et al.  2007; Ferguson et al. 2005; Kalirai et
al. 2006), and  beyond the Local Group (e.g.,  Malin \& Hadley 1997;
Forbes et al. 2003;  Pohlen et al.  2004; Martinez-Delgado \etal 2012).
In additional, the unbound stars  from satellite galaxy are thought to
be the origin of the diffuse intracluster light in clusters (Arnaboldi
et al. 2002; Gerhard et al.   2005; Mihos et al. 2005; Gonzalez et al.
2000; Zibetti et al. 2005),  and even in galactic stellar halo (Bullock
\& Johston  2005; Bell  \etal 2008; Carollo  \etal 2010;  Cooper \etal
2010).

These galaxy interactions can also lead  to important morphological
transformations in the satellite stellar  distribution.  Barnes (1992)
showed that a  stellar disc can be destroyed and  form a hot, pressure
supported spheroidal galaxy through tidal heating and violent relaxation
during   mergers.    Moreover   there   is  mounting   evidence   that
environmental effects  (e.g.  tides  and stripping) have  an important
role in shaping the properties  of the local dwarf spheroidal galaxies
(e.g., Einasto  et al. 1974;  Faber \& Lin  1983; Mayer et  al. 2001a,
2001b; Kravtsov et al. 2004;  Mayer et al. 2006, 2007; Klimentowski et
al. 2007; Penarrubia et al. 2008; Klimentowski et al. 2009a, 2009b).

The  above  two process  ,  tidal  stripping  of satellite  stars  and
morphology  change in  satellite remnant,  have great  impacts  on the
evolution of  both satellite and  central galaxy.  For example,  it is
observationally found  that massive central galaxies  grow little in mass from
$z=1$ to $z=0$ (e.g., Fontana et  al.  2004), this seems to be in contrast
with predictions based on the CDM model, where a large number
of satellites should be accreated in this time interval.
This observational result could be accounted for 
in galaxy formation models only if a considerable amount of the satellite 
stellar component is removed during the orbit, before the merger actually happens.
(e.g.,  Monaco et al. 2006; Somerville et al. 2008;  Kang et al. 2008).  
Moreover knowing the mass (stellar and dark matter) of the satellite before the merger
is crucial to correctly model the effects of the merger itself, like remnant morphology, 
induced star formation etc. (e.g, Kauffmann et al. 1999; Cole et al. 2000).
Thus a  full understanding  of the evolution  of satellite  galaxy, in
term  of their mass removal  and morphology  change, is  important to
correctly model  galaxy formation.   

A  number of  studies  have used  numerical
N-body  simulations to  investigate the  influence of  environments on
satellite evolution, both  on cluster mass scales (e.g.,  Moore et al.
1999; Gnedin 2003;  Mastropietro et al.  2005), and  on Milky Way-like
mass scales (e.g.,  Mayer et al.  2001a,b; Klimentowski  et al.  2009a;
Kazantzidis  et al.   2011).   Recently Villalobos  \etal (2012)  have
presented a very thorough study,  covering a broad parameter space, on
the  effects  of  environment  on  the  satellite  galaxies  in  group
environment.   
 
In this paper we use N-body simulations to study the effect 
of the initial satellite and central galaxy morphology  on the efficiency 
of tidal stripping and mass loss. 
We   adopt  three  different morphologies for the satellite: 
pure disc, pure bulge and the combined one with  80\% of stellar mass in the disc.
We also employ two morphologies for the central galaxy, namely pure disc and the combined bulge+disc.
We vary the orbit of the satellite, changing the angle of the orbital plane with
respect to the central disc, and we test both prograde and retrograde orbits.  
The paper is organized as  follows.  \S \ref{sec:setup}
describes  the  numerical methods  and  the  parameters  we used.   \S
\ref{sec:results}  shows the  results for  different satellite
morphologies,  different   host  morphologies,  and   different  orbit
parameters.   Conclusions  and  discussion are  presented  in  \S
\ref{sec:conclusions}.

\section{Numerical methods}
\label{sec:setup}

The parameter space describing galaxy mergers is very wide, since it 
includes several structural parameters defining the merging galaxies as 
well as the orbital parameters of the merger itself.
In this work we have decided to concentrate our attention on the effects
of morphology on stellar and dark matter mass removal.
We will then keep the mass (stellar and dark matter) of the host and the
satellite fixed through this paper, while we will vary how the stellar
mass is distributed within both galaxies.
The initial conditions are created using the same approach as in Springel 
et al. (2005), the simulations are evolved with the {\sc gadget2} code
(Springel et al. 2005) and we use {\sc subfind} to identify the bound structures.
Here below more details on all those three steps are provided.

\subsection{Initial Conditions}

Each galaxy is composite of a dark matter halo and a stellar component. 
The dark matter halo is modeled with a Hernquist density profile (Hernquist 1990):
\begin{equation} \label{eq:Hernquist}
\rho_{\rm{dm}} (r)=\frac{M_{\rm{dm}}}{2\pi}\frac{a}{r (r+a)^3} ,
\end{equation}
where $a$  is the radial scale length.  We require that this density profile
to be the same of a NFW one (Navarro, Frenk and White 1997) with the same total mass.
Under this requirement there is a one to one correlation between the $a$ parameter
and the scale radius ($r_s$) of the NFW profile:
\begin{equation} \label{eq:a-rs}
a=r_{\rm s}\sqrt{2[\ln (1+c)-c/ (1+c)]} ,
\end{equation}
where $c$ is the concentration parameter, defined as $c=r_{200}/r_{\rm s}$,
and $r_{200}$  is the radius at  which the mean enclosed
dark matter  density is 200 times  the critical density.

We tight the halo concentration to the halo mass using the well
known concentration-halo mass relation, with the parameters given 
in Macci\`o et al. (2008). We fix the mass of the primary halo to be
$10^{12} \Msun$ and for the satellite halo, we used a mass ratio of 1/8, 1/16
and 1/24. 
The spin of the dark matter halo is fixed to be $\lambda=0.03$ for 
both halo masses, in good agreement with results from cosmological 
simulations (e.g. Macci\`o et al. 2007).

The stellar  mass of  each galaxy is  set using the  halo-stellar mass
relation from Moster  et al.  (2010), based on the abundance matching techniques.
As a conquense, the mass ratios of stellar mass between perimal and satellite
are 1:24, 1:120 and 1:240, respectively. Table \ref{tab:galaxy-param} contains detailed lists of all parameters
we kept fixed in all our runs. While Table \ref{tab:lightersat} lists the
varable ones.

For the distribution of the stellar component in each galaxy 
we employ three different morphologies:

\begin{itemize}
\item A rotationally supported disc galaxy (D). The stellar  disc has an
  exponential  surface density  profile  with scalelength  $R_s$, and 
  Spitzer's   isothermal   sheet    with   scaleheight   $z_0$.   The
  three-dimensional (3D) stellar density in the disc is hence given by
\begin{equation}\label{eq:disc}
\rho_{\star} (r, z)=\frac{M_{\star}}{4\pi z_0R_s^2} \mathrm{sech} ^2\left
(\frac{z}{2z_0}\right)\exp\left (-\frac{R}{R_s}\right)
\end{equation}
\item  A non-rotating  spheroidal  bulge galaxy (B). 
  The stellar bulge has a Hernquist density profile (see eq.~\ref{eq:Hernquist}).
\item A composite  galaxy with both Hernquist bulge  and an exponential disc  (C). 
  The  bulge contains 20\% of the total stellar mass unless stated otherwise.
\end{itemize}

From now on, we will use the capital letter ``C'' and ``B''to describe
the morphology of the primary  (host) galaxy, and use ``c'', ``b'' and
``d'' for the satellite. For example, an C-d merger will indicate 
a the  merger  between a composite primary galaxy (disc+bulge) and a pure disc satellite galaxy.
More  details   can  be   found  from Table.~\ref{tab:morph-param}.

The disc scalelength $R_s$ is set by relating it to the angular momentum
$J_d$ of the disc, assuming strict centrifugal support of the disc and
negligible disc thickness compared with its scalelength, so we have:
\begin{equation}\label{eq:jandh}
J_d=j_dJ=M_d\int_0^{\infty}V_c (R)\left (\frac{R}{R_s}\right)\exp\left (-\frac{R}{R_s}\right)\ud R
\end{equation}
where $j_d= (M_{disk}/M_{200})$ by  assuming that  the disc  has the  same specific
angular momentum as  the dark matter halo. The  disc scaleheight $z_0$
is  set as  $z_0=0.2R_s$. A complete list of all parameters describing the stellar component
can be found in Table~\ref{tab:morph-param}.

\begin{table}
\centering
\begin{minipage}{80mm}
 \caption{Properties of the host and reference satellite galaxies}
 \label{tab:galaxy-param}
 \begin{tabular}{@{}lllr@{}}
  \hline
                                     & ``Primary''    & ``Satellite''            &               \\
  \hline
  \hline
  DM Halo                            &                &                      &               \\
  \hline
  Virial mass ($M_{200}$)            & 96       & 11   &  ($10^{10}M_{\sun}$)  \\
  Virial radius ($r_{200}$)          & 160      & 80   &  (kpc)         \\
  Concentration ($c$)                & 6.8      & 8.3  &               \\
  spin parameter ($\lambda$)         & 0.03     & 0.03 &               \\
  Number of particles                & 8        & 8    &  ($10^5$)             \\
  Softening  ($\epsilon_h$)          & 0.3      & 0.1  &  (kpc)         \\
  \hline
  Stellar (disc \& bulge)            &          &      &               \\
  \hline
  Mass ($M_{\star}$)$^{(1)}$         & $3\%$    & $1\%$ &  ($M_{200}$)             \\
  Number of particles                & 6        & 2     &  ($10^5$)             \\
  Softening ($\epsilon_s$)           & 0.07     &  0.02 &  (kpc)         \\
  \hline
  \hline
 \end{tabular}
\\
 (1):$M_{\star}$ for the stellar mass in unit of the virial mass of dark matter
 halo $(M_{200})$.
\end{minipage}
\end{table}

\begin{table*}
\centering
\begin{minipage}{160mm}
 \caption{Stellar morphologies of hosts and reference satellite}
 \label{tab:morph-param}
 \begin{tabular}{@{}l | llllllllllr@{}}
  \hline
                              & \multicolumn{2}{c}{``Primary''} & \multicolumn{7}{c}{ ``Satellite''}& \\
  
  \hline
  Morphology                  & BULGE & COMPOSED & \multicolumn{2}{c}{bulge} &
  \multicolumn{4}{c}{composite} & \multicolumn{2}{c}{disc} & \\
  \hline
                              & B & C & b & bL$^{(8)}$ & cB$^{
				      (9)}$ & c & cT$^{ (10)}$ & cD$^{ (11)}$ &
				      d & dT$^{ (10)}$ & \\
  \hline
  $M_{disc}$$^{(1)}$       & 0   & 0.8 & 0   & 0    & 0.7  &0.8  & 0.8  & 0.9  &
  1.0 & 1.0  &  ($M_{\star}$) \\
  $j_d$$^{(2)}$  & 0   & 0.024 & 0 & 0    & 0.007&0.008&0.005 & 0.009& 0.01& 0.006&              \\
  $R_s$$^{(3)}$    & 0   & 3.01 & 0  & 0    &1.59 &1.59 & 1.04 & 1.59 & 1.60& 1.01 &  (kpc)        \\
  $N_{disc}$$^{(4)}$   & 0   & 4.8 & 0   & 0    &1.4  & 1.6 & 1.6  & 1.8  & 2.0 & 2.0  &  ($10^5$)     \\
  \hline
  $M_{bulge}$$^{(5)}$     & 1.0 & 0.2 & 1.0 & 1.0  &0.3  & 0.2 & 0.2  & 0.1  & 0
  & 0    &  ($M_{\star}$) \\
  $a^{(6)}$     & 0.61& 0.62& 0.29& 0.72 &0.32 & 0.32& 0.32 & 0.21 & 0   & 0    &  (kpc)        \\
  $N_{bulge}$$^{(7)}$   & 6.0 & 1.2 & 2.0 & 2.0  &0.6  & 0.4 & 0.4  & 0.2  & 0   & 0    &  ($10^5$)     \\
  \hline
  \hline
 \end{tabular}
\\
 (1):$M_{disc}$ for the stellar disc mass in unit of total stellar mass
 $M_{\star}$.
 (2):$j_d$ for the disc spin parameter.
 (3):$R_s$ for disc scale length.
 (4):$N_{disc}$ for the particles number of disc.
 (5):$M_{bulge}$ for the stellar bulge mass in unit of total stellar mass
 $M_{\star}$.
 (6):$a$ for bulge scale length.
 (7):$N_{bulge}$ for the particles number of bulge.
 (8):``L'' for Larger with $a=0.72kpc$. 
 (9):``B'' for more Bulge with mass fraction $M_{bulge}=0.3M_{star}$.
 (10):``T'' for Thicker disc than reference disc  (see $R_s$).
 (11):``D'' for more Disk with disc fraction $M_{disc}=0.9M_{star}$.
\end{minipage}
\end{table*}

\begin{table}
\centering
\begin{minipage}{75mm}
 \caption{Properties of lighter satellites}
 \label{tab:lightersat}
 \begin{tabular}{ llllr}
  \hline

            & bulge        & \multicolumn{2}{c}{disk} &   \\
  \hline
  \hline
            & b$_5^{(1)}$ & d$_5$   & d$_{10}$     &   \\
  \hline
  $M_{200}$ & 6.0          & 6.0      & 4.2           & ($10^{10}M_{\sun}$)  \\
  $r_{200}$ & 63.48        & 63.48    & 56.65         & (kpc) \\
  c         & 8.91         & 8.91     & 9.22          &       \\
  $\lambda$ & 0.03         & 0.03     & 0.03          &       \\
  Nhalo     & 5            & 5        & 4             & ($10^5$) \\
  $\epsilon_h$ & 0.2       & 0.2      & 0.2           & (kpc)  \\
  \hline
  $M_{\star}$ & 0.37\%     & 0.37\%   & 0.26\%        & ($M_{200}$) \\
  $R_s$     & 0.23         & 1.35     & 1.12          & (kpc)  \\ 
  Nstar     & 0.4          & 0.4      & 0.2           & ($10^5$) \\
  $\epsilon_s$ & 0.05      & 0.05     & 0.05          & (kpc)   \\
  \hline \hline
 \end{tabular}
 \\
 (1): ``b$_5$'' for the bulge satellite whose stellar component is 5 times less massive than the
 reference satellite and ``d$_{10}$'' for the disk satellite with 10 times less
 massive on stellar component comparing with the reference satellite. 
\end{minipage}
\end{table}

\subsection{Orbital parameters}

For each experiment, the satellite galaxy is initially released at the
virial   radius  of  the   primary  galaxy   with  velocity   $  (V_r,
V_{\varphi})$,           where           $V_r=v_rV_{200}$          and
$V_{\varphi}=v_{\varphi}V_{200}$   are  the   radial   and  tangential
components  of the  initial velocity,  respectively. $V_{200}$  is the
circular  velocity  at  the  virial  radius  of  the  central  primary
galaxy. Our reference orbit has been set with $(v_r, v_{\varphi})=  (0.9, 0.6)$
as suggested by results based on cosmological simualations (Benson 2005).
We also explore more ``circular''  or  ``radial''  orbits  by changing $v_r$ and $v_{\varphi}$.

\begin{figure}
\includegraphics[width=75mm]{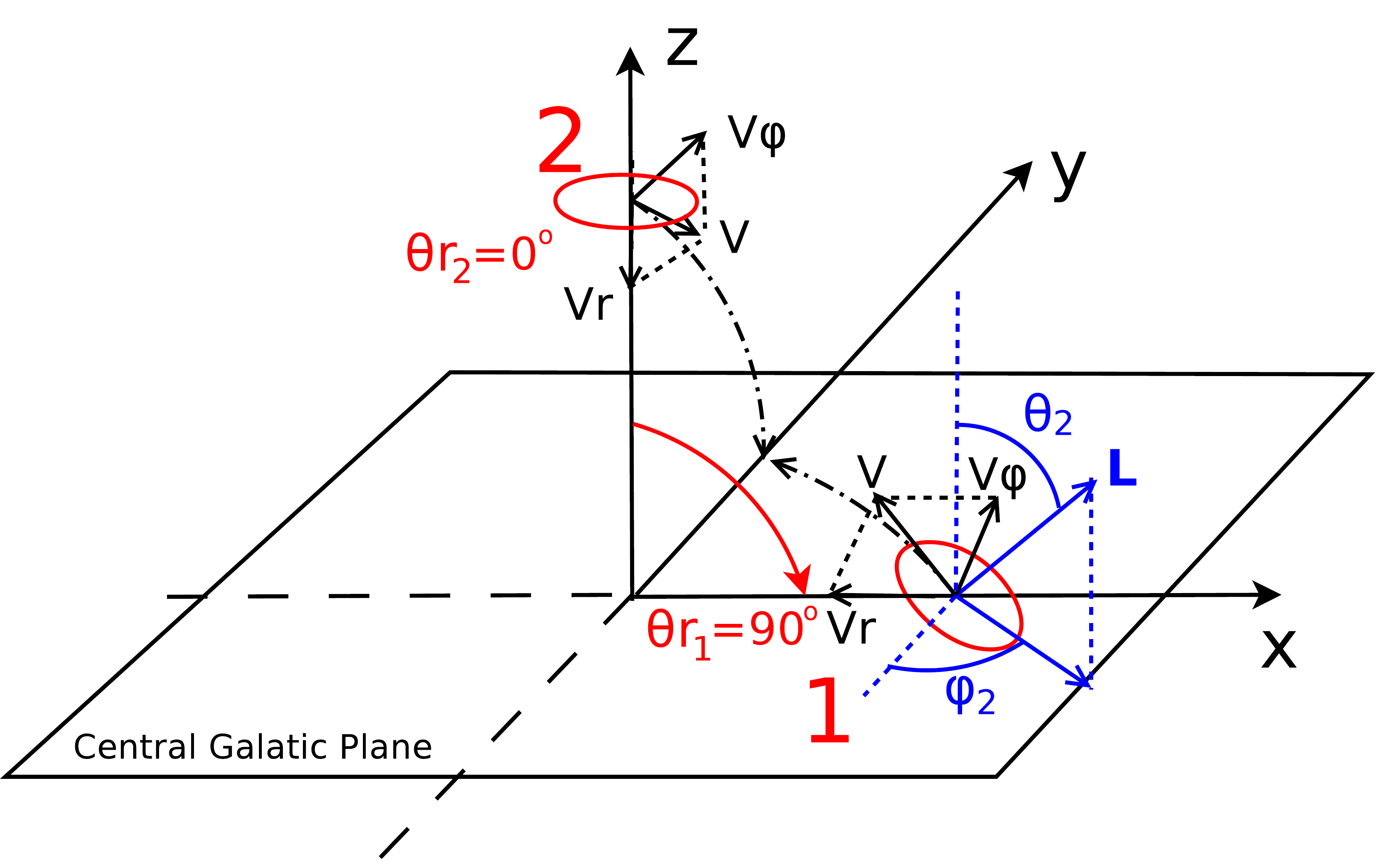}
\caption{Initial  satellite galaxy  position and  velocity  in the primary
  galaxy frame. $V_r$ and  $V_{\varphi}$ are the radial and tangential
  components  of  the  initial  velocity, respectively. Case 1 represents a
  coplanar orbit with respect to the primary galaxy disc, with the
  satellite polar angle $\theta_{r_1}=90^{\circ}$. {\bf L} indicates the angular
  momentum of stellar disc. $\theta_2$ \& $\varphi_2$ are the azimuthal and
  polar angle of {\bf L}. Case 2 represents orbit
  perpendicular with $\theta_{r_2}=0^{\circ}$.}
\label{fig:orbit}
\end{figure}

If disc galaxies are used for the central and/or the satellite, there are further
degrees for freedom that are related to the angles between the two discs and
between the satellite disc and the orbital plane.
These three angles are described in Figure~\ref{fig:orbit}.

A detailed list of all the orbital parameter  used in the different merging
simulations can be found in Table~\ref{tab:orbit-param}.

\begin{table}
\centering
\begin{minipage}{75mm}
 \caption{Orbit parameters.}
 \label{tab:orbit-param}
 \begin{tabular}{@{}lllllll@{}}
  \hline
Label   & Host & sat. &  ($v_r, v_{\varphi}$) & $\theta_2$ & $\varphi_2$ & $\theta_r$ \\ 
  \hline
  \hline
C-c     & C  & c    &  (0.9,0.6)          & 0          & 0           & $90^\circ$  \\
C-cR$^{ (1)}$& C & c  &  (0.9,0.6)          & 180        & 0           & $90^\circ$  \\
C-cB    & C  & cB   &  (0.9,0.6)          & 0          & 0           & $90^\circ$  \\
C-cD    & C  & cD   &  (0.9,0.6)          & 0          & 0           & $90^\circ$  \\
C-cT    & C  & cT   &  (0.9,0.6)          & 0          & 0           & $90^\circ$  \\
\hline
C-d     & C  & d    &  (0.9,0.6)          & 0          & 0           & $90^\circ$  \\
C-dT    & C  & dT   &  (0.9,0.6)          & 0          & 0           & $90^\circ$  \\
C-dR    & C  & d  &  (0.9,0.6)          & 180        & 0           & $90^\circ$  \\
C-dTR   & C  & dT   &  (0.9,0.6)          & 180        & 0           & $90^\circ$  \\
C-dc$^{(2)}$& C& d  &  (0.6,1.1)          & 0          & 0           & $90^\circ$  \\
C-dr$^{(3)}$& C& d  &  (1.2,0.3)          & 0          & 0           & $90^\circ$  \\
C-d45   & C  & d    &  (0.9,0.6)          & 0          & 0           & $45^\circ$ \\
C-d90   & C  & d    &  (0.9,0.6)          & 0          & 0           & $0^\circ$ \\
\hline
C-b     & C  & b    &  (0.9,0.6)          & 0          & 0           & $90^\circ$  \\
C-bL    & C  & bL   &  (0.9,0.6)          & 0          & 0           & $90^\circ$  \\
C-b45   & C  & b    &  (0.9,0.6)          & 0          & 0           & $45^\circ$ \\
C-b90   & C  & b    &  (0.9,0.6)          & 0          & 0           & $0^\circ$ \\
\hline
C-d$_{10}$& C & d$_{10}$& (0.9,0.3)       & 0          & 0           & $90^\circ$ \\
C-b$_5$& C  & b$_5$& (0.9,0.3)            & 0          & 0           & $90^\circ$ \\
C-d$_5^1$& C & d$_5$& (0.9,0.3)           & 0          & 0           & $90^\circ$ \\
C-d$_5^2$& C & d$_5$& (0.8,0.5)           & 0          & 0           & $90^\circ$ \\
C-d$_5^3$& C & d$_5$& (0.9,0.4)           & 0          & 0           & $90^\circ$ \\
\hline
B-c     & B  & c    &  (0.9,0.6)          & 0          & 0           & $90^\circ$  \\
B-d     & B  & d    &  (0.9,0.6)          & 0          & 0           & $90^\circ$  \\
B-b     & B  & b    &  (0.9,0.6)          & 0          & 0           & $90^\circ$  \\
\hline
C-dv0$^{ (4)}$& C& d &  (0.9,0.6)          & 90         & 0           & $90^\circ$  \\
C-dv90  & C  & d    &  (0.9,0.6)          & 90         & 90          & $90^\circ$  \\
C-dv180 & C  & d    &  (0.9,0.6)          & 90         & 180         & $90^\circ$  \\
C-dv270 & C  & d    &  (0.9,0.6)          & 90         & 270         & $90^\circ$  \\
  \hline
  \hline
  \end{tabular}
\\ (1):``R'' for Retrograde encounter, the same below.
 (2):``c'' for more ``circular'' encounter with orbital eccentricity $e=0.6$.
 (3):``r'' for more ``radial'' encounter with $e=0.97$. 
 (4):``v'' for ``vertical'' disc with the polar angle of angular momentum of the
 disc $\theta_2=90^\circ$ 
respected to the orbit plane. 
\end{minipage}
\end{table}

\subsection{Bound and unbound mass of the satellite galaxy}
\label{ssec:bound}

Here  we describe  our procedure to find  the bound  mass of the  satellite
galaxy along its orbit. 
At  each time step we: (i) pick  up all the stellar  particles of the
satellite  by  their  IDs,  and  get the  position  of  the  highest
density point. (ii) Get the mean velocity  of all particles within the radii
$R_{core}$ ( ``core'') around  the density peak, and  it is taken as
the  background velocity.   (iii)  Compute  the gravitation  potential
energy $E$ and kinetic energy $K$ of each particles, for both star and
dark  matter, in  the  core's frame,  and if  $E<K$,  the particle  is
unbound, and removed from the  remnant.  (iv) Repeat the previous step
until the  remnant has converged.  We assume  that if the  remnant has
stellar   particles   less  than   25,   the   satellite  is   totally
disrupted. Those  unbound particles  are stripped  by the  tidal force
from the host galaxy.

The only  free parameter in the  above procedure is $R_{cor}$  used to
define   the    ``core''   size.   In   our    simulation   we   adopt
$R_{core}=0.1kpc$, and  we found  that our results  are stable  to the
slightly changing of $R_{core}$ by a  few factor. A too large or small
$R_{core}$ will result in large fluctuation on the remnant mass during
the satellite evolution.

\section{Results}
\label{sec:results}
\subsection{Effects of satellite morphology}

\begin{figure*}
\includegraphics[width=155mm]{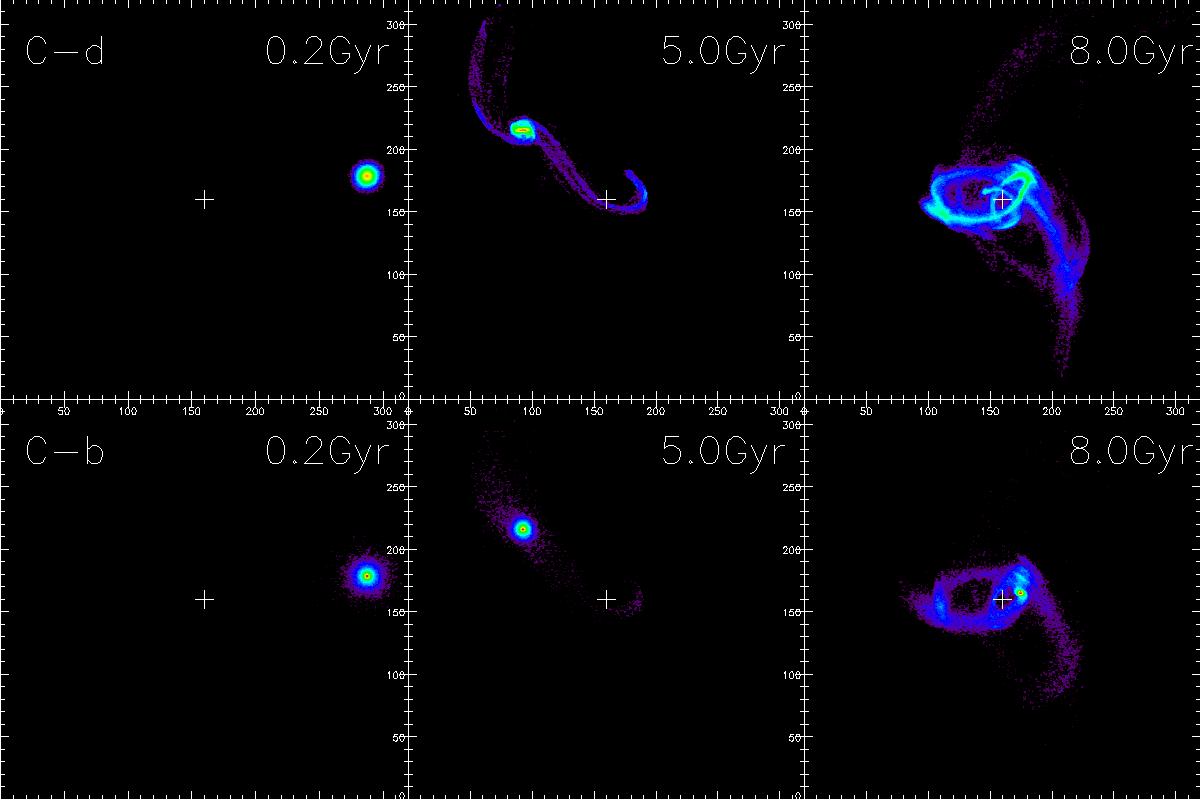}
\caption{Time  evolution  of  the stellar component of satellite galaxy. The
cross indicates the center
  of  the primary  galaxy.   The  upper  panels  shows  the
  C-d merger, while the lower one is the C-b merger. The unit of  the coordinate
  is kpc, and the color shows the density.}
\label{fig:showstar}
\end{figure*}

\begin{figure*}
\includegraphics[width=135mm]{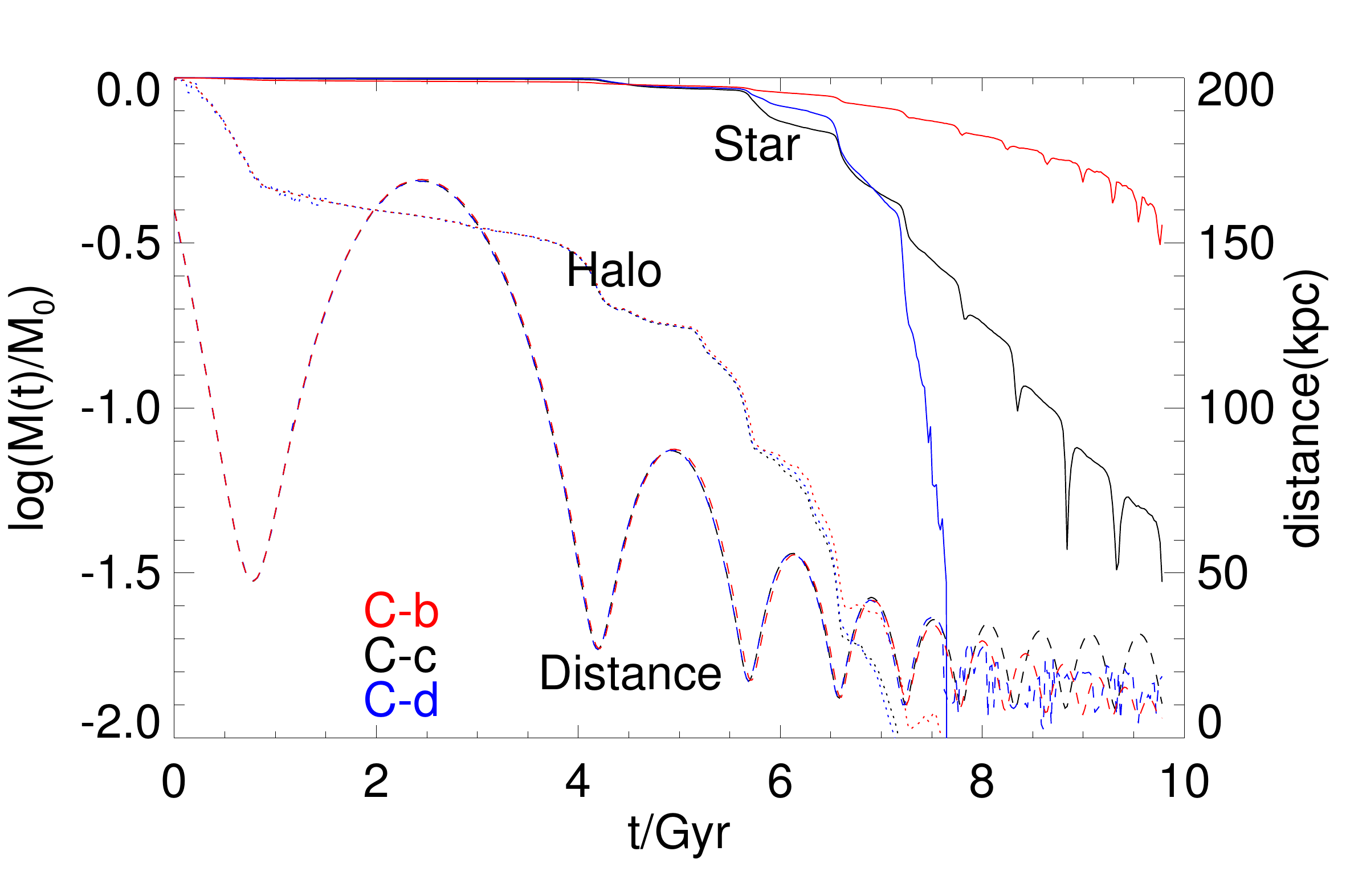}
\caption{The distance and remnant bound mass fraction as
  a function of  time for 3 different satellite morphologies
  merging with a bulge + disc host.
  The solid lines show the evolution of
  stellar mass, while  the  dotted shows one of the dark
  matter component.  The dashed lines indicate the distances  between satellite
  and center host galaxy.}
\label{fig:dislogbind3}
\end{figure*}

In this section we analyze there different mergers where the center galaxy is
kept fix (Composite) and we vary the morphology of the incoming satellite (disc, bulge
and composite). The upper pannels of Fig.~\ref{fig:showstar} show the evolution
of the stellar mass density for a pure disc (d), while the lower
ones for satellite with pure bulge (b).
The primary galaxy is always at the center of each panel (labeled with a cross)
and its stellar component is not shown.

The  upper  panels show  that, at beginning of  galaxy approach
(left panel, t=0.2 Gyr),  the satellite  still remains its
initial disc shape (the disc is face-on in this projection).
A prominent tidal feature  appears soon after
the second  pericentric passage (t=5 Gyrs).  
Such tidal  features are  very common
during the galaxy interactions (e.g., Toomre  \& Toomre 1972; 
Dubinski et al.  1996). Much more  twisted tidal  tails can  be seen  
after a  few pericentric passages (the upper right, t=8.0 Gyrs). 
The galaxies then is completely disrupted at t=8.5 Gyrs.

The  situation  is different  in  the  lower series of panels
where the evolution of a pure bulge satellite is shown.
After  the  second pericentric  passage
(t=5.0 Gyrs) the  tidal features are very weak, they become then more
evident at t=8.0 Gyrs (third panel), but even in this case the satellite
still retains more than 80\% of its initial stellar mass.
Finally the satellite is able to survive to tidal force for the whole
simulation time (10 Gyrs) with a total stellar mass loss of less than 40\%.

By simply looking at Figure~\ref{fig:showstar} it is already evident that 
satellite morphology plays a key role in shaping stellar stripping.

We quantify these different behaviors in Fig.\ref{fig:dislogbind3}, 
where the evolution of the stellar mass (solid lines), halo mass (dotted lines)
and distance from the center (dashed lines) is shown for the three
different satellite morphologies (keeping fixed the morphology of the central 
galaxy, C).

The comparison of the solid with the dotted lines show that the stripping of the dark matter
halo  is  quite  different  from  that of  the  stars.  At  the  first
pericentric passage, the halo has lost  its mass by around 50\%, while
stars are practically untouched. The removal of the stellar component starts only after having 
lost 80-90\% of the dark matter mass. The stellar mass loss
is always slower than the dark matter one, and the satellites becomes
more and more stellar mass dominated with time (we will come back 
on this issue later).
The above  results are  very similar  to the  recent ones  obtained by
Villalobos et al. (2012).  These authors run a series of simulation of
satellite orbiting in a ``group'' environment. They also find that the
satellite DM halo loses mass quickly than the disc, and after the first
pericentric  passage, the  halo has  lost around  50\% of  its initial
mass. 

Let's now look at the stellar mass loss. At the initial stage of 
the merger the stellar mass is somehow
shielded by the dark matter halo and no stripping is present for
$\approx 5$ Gyrs. After the third pericentric passage the stellar
mass loss begins. The C-d and the C-c merger show a similar mass
loss for 1.5 Gyrs, while the b satellite is able to retain a larger
fraction of its stellar mass. Things change dramatically at $t\approx 7.2$ Gyrs, 
when the mass of the satellite in the C-d scenario drops by more
than two orders of magnitude in less than 200 Myrs, with the satellite
being practically destroyed at $t\approx 8$ Gyrs.

The C-b run has a complete different behavior, with the satellite
able to retain more than 60\% of its mass at the end of the
simulation run ($t=10$ Gyrs). The C-c case is somehow in the middle;
it experiences a strong mass loss in the initial phase (when the disc
is removed), but then is able to retain up to 20\% of its initial
bulge mass till $t=10$ Gyrs. Our results are in fair good agreement with
Villalobos \etal(2012). One difference is for the composite (bulge + disc) case
where Villalobos \etal found a faster stripping with respect to a pure disc
case. The reason for this discrepancy lies in the fact that Villalobos \etal
``added'' the bulge to the disc hence changing the total stellar mass. While in
our case $M_{\star}$ is a constant. A larger mass implies a different orbital
evolution and then a different mass loss rate.
From Fig.\ref{fig:dislogbind3} it is clear that the satellite morphology has a very strong
impact on its tidal disruption. Two satellites with the same stellar and dark matter
mass, on the same orbit could either retain 60\% of their mass after 10 Gyrs or being
completely destroyed only depending on the initial stellar morphology.

Finally, it is interesting to note that the fate of the stripped material
is independent from satellite morphology. In all cases  after 10 Gyr, around
15\% of unbound stellar mass of satellite, no matter disk and bulge, locates
within 10 kpc away from host center. The other unbound component exists as
diffuse component in host galaxy halo, which could be the origin of the
intra-galactic light.

\subsection {Tidal radius and disc scale length}

\begin{figure*}
\includegraphics[width=155mm]{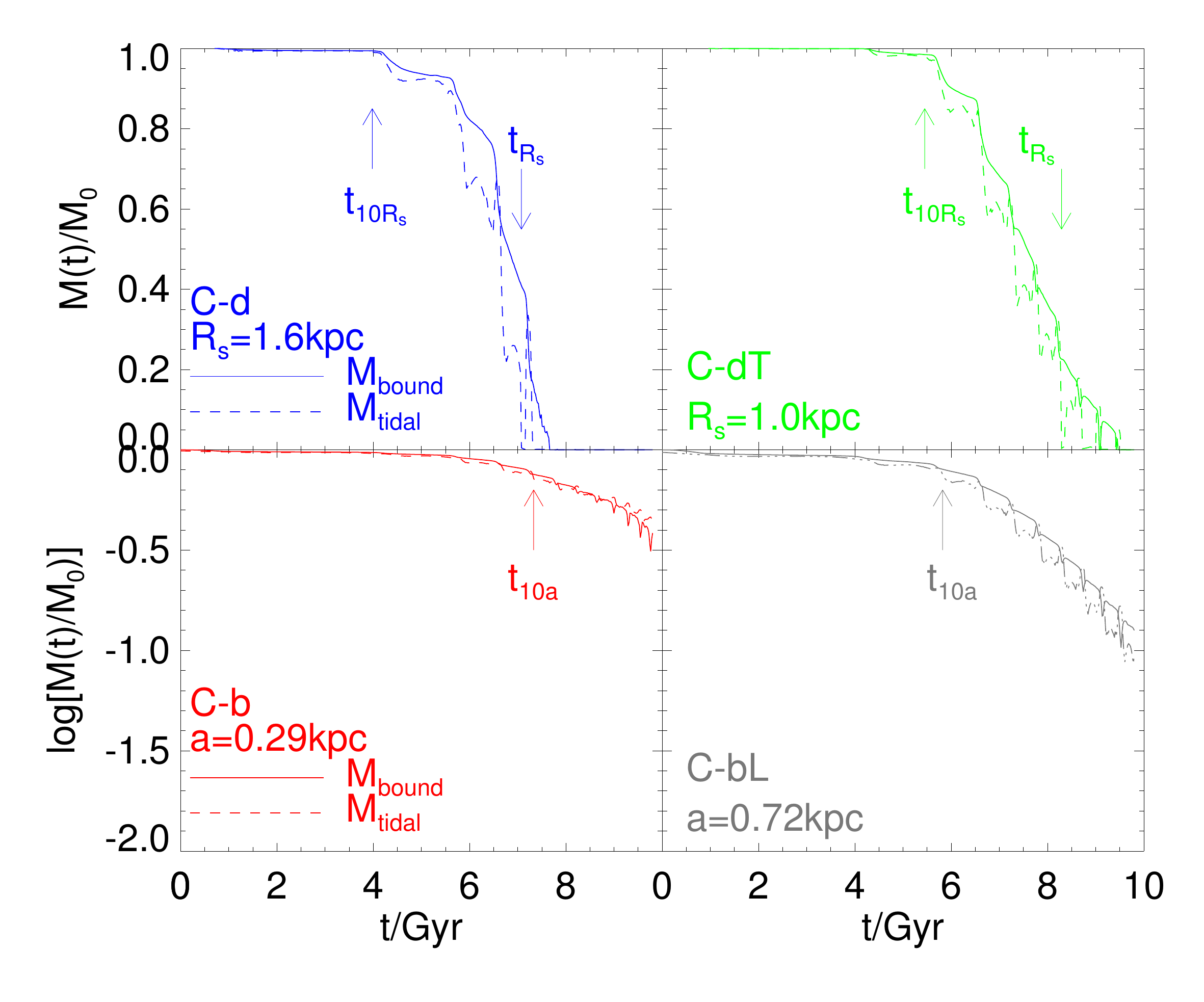}
\caption{The correlation  between tidal mass  and bound mass.  The left
  upper panel shows the evolution of remnant bound mass $M_{bound}$
  (solid line)  and the mass  within tidal radius  $M_{tidal}$ (dashed
  line)  of  a  disc  satellite.   The upward  arrow  shows  the  time
  $t_{10R_s}$  when $R_{tidal}  (t_{10R_s})=10R_s$,  and  the downward
  arrow shows $t_{R_s}$  when  $R_{tidal}  (t_{R_s})=R_s$.  The  upper right
  panel shows the same thing but for a thick disc satellite with $R_s=1.0kpc$.
  Bottom panels show the evolution for bulged satellite for different values of
  a (left a=0.29, right a = 0.72).}
\label{fig:bindtidalmgdb}
\end{figure*}

We want now to look deeper into the time evolution of 
the disc mass in the C-d merger.
Our goal is to understand what sets the specific moment in time when the disc
experiences the exponential stellar mass loss shown in 
Fig.\ref{fig:dislogbind3}. 
As a first thing we compute the tidal radius for satellites.
We define the tidal radius ($R_t$) according 
to Binney \& Tremaine (2008) as the distance between the Lagrange point, $L_3$,
and the center of the satellite. We then define the tidal mass ($M_t$) as the 
mass within $R_t$.
With this definition we find a good agreement between the tidal mass
and the bound mass (see section \ref{ssec:bound}), as shown by 
the dashed and the solid lines in the four panel of Fig.~\ref{fig:bindtidalmgdb}.
The slightly offset between the two curves is due to non-instantaneous
removal of the mass outside $R_t$, as previously found by in other
studies (e.g., Taylor \& Babul 2001; Zentner  \& Bullock  2003; Gan  et  al. 2010).

We found two characteristic time scales: the first one determines the beginning 
of the stripping effect and it is set by the time when the tidal radius is equal
to 10 times the disc scale length ($t_{10R_s}$), the second one marks the moment
when the stellar tidal stripping becomes exponential, this happens when the tidal radius
is of the same order of the disc scale length ($t_{R_s}$).
These two time scales are marked by vertical arrows in the upper left panel 
of  Fig.~\ref{fig:bindtidalmgdb}.

In order to test whether these time scales are really independent from the 
actual value of $R_s$, we run second simulation with a thicker disc (C-dT) with 
a smaller scale length. Results are shown in the right upper panel of
Fig.~\ref{fig:bindtidalmgdb}, where we find that $t_{10R_s}$ and $t_{R_s}$ 
are still able to capture the transition between different stripping regimes
(this also applies for runs with different orbital parameters
see Fig.~\ref{fig:massratio}.)

According to our results there are two time scales for disc galaxies
the first one  ($\uptau_{life}$) defines the life time of satellite
stellar disc, the second one ($\uptau_{strip}$) defines the time
from the beginning of the stellar tidal stripping  to the complete
disruption of the satellite, in formulas:
\begin{equation}
\uptau_{life}=t_{R_s}-t_{acc}\\
\uptau_{strip}=t_{R_s}-t_{10R_s}.
\end{equation}
Where $t_{acc}$ is the time of accretion of the satellite onto the main halo.
Note  that our  definition of  $\uptau_{life}$ is  different from  the
conventional ones,  such as the total satellite dynamical friction time-scale.
This time scale should  be  incorporated in models of  galaxy formation, to
predict the stellar mass loss for  the disc of satellite galaxy. 

On the other hand we do not find a specific time scale
for the bulge disruption (as shown in lower panels of
Fig.~\ref{fig:bindtidalmgdb} for two simulations with
different bulge scale lengths). It is worth noting that, for the same stellar
mass, the larger the bulge is the less amount of mass it retains at the end, as
expected for a less compact object. The bulge mass removal
seems to have a constant efficiency that we expect to 
be orbital dependent, which will be discussed in \S~\ref{sec:sat_orbit}.

\subsection{Satellite mass profile and morphological evolution}

\begin{figure*}
\includegraphics[width=155mm]{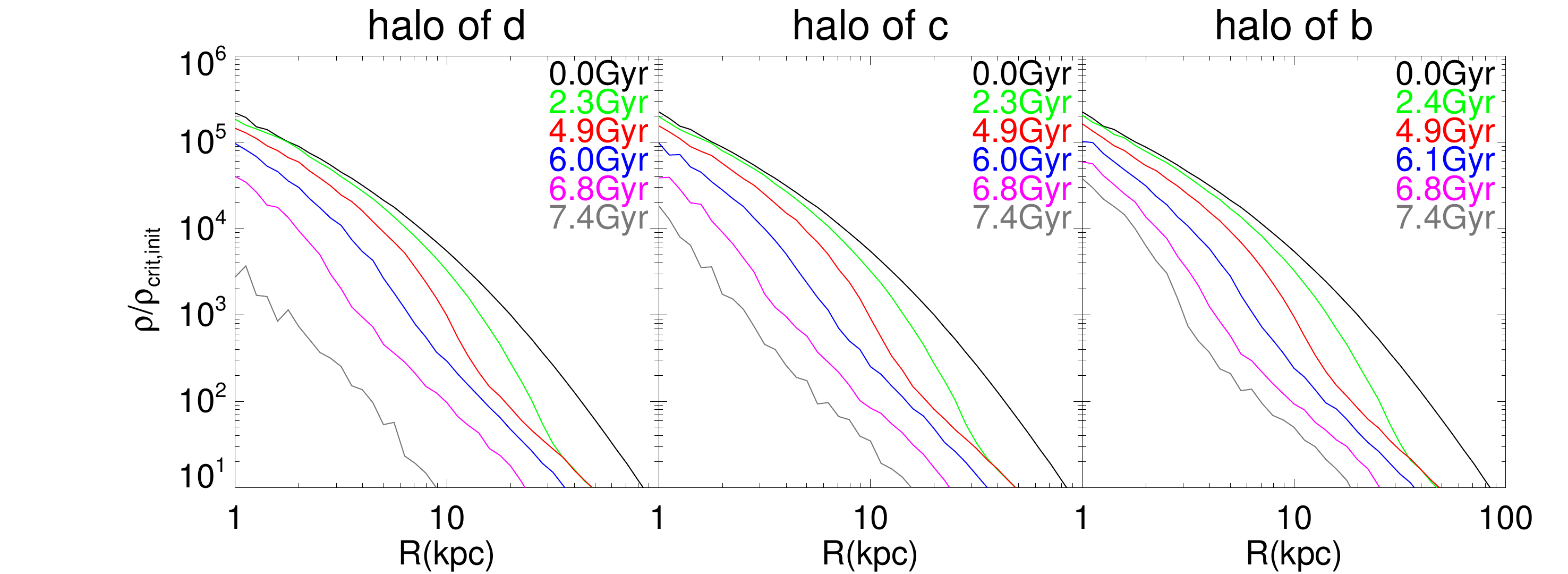}
\caption{The  evolution  of the dark matter density profile for three different
satellite morphologies. From left to right are ``d'', ``c'' and ``b''. The density profiles
are shown at different apocentric passages.}
\label{fig:halodenspro}
\end{figure*}

\begin{figure*}
\includegraphics[width=155mm]{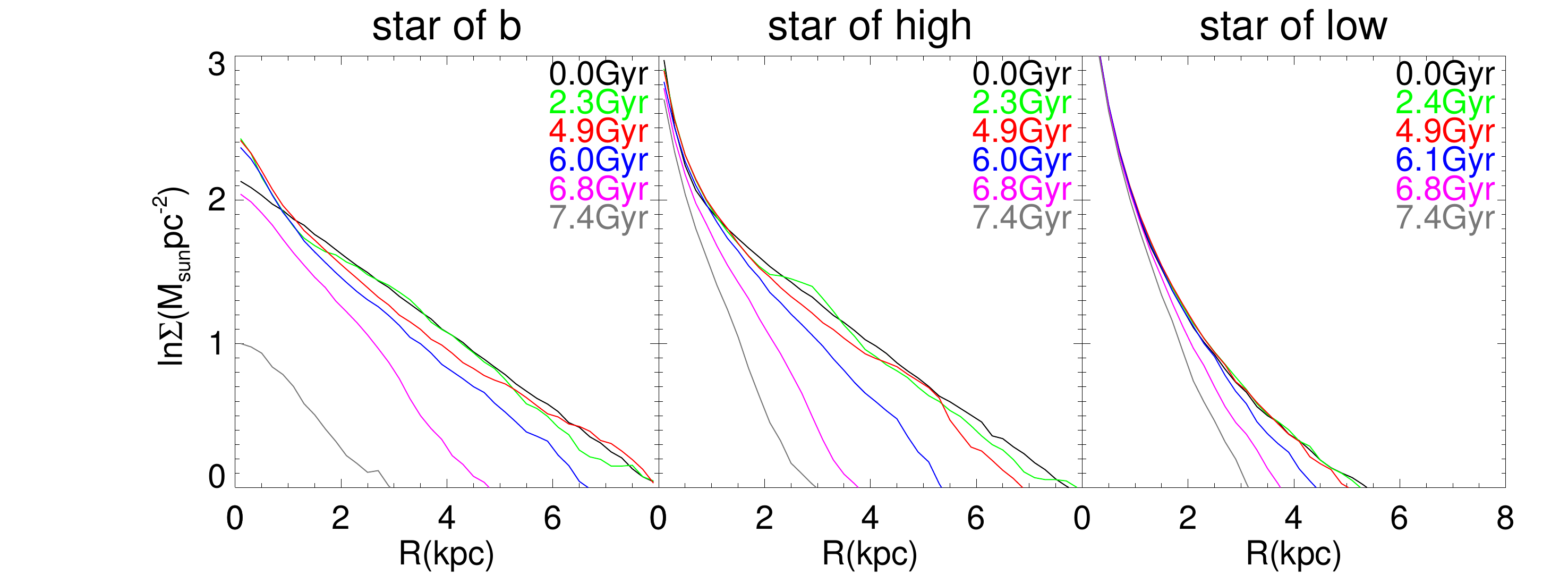}
\caption{The evolution of the surface density profile of stellar component for
three different satellite morphologies. From left to right are ``d'', ``c'' and
``b''. Also the density profiles are shown at different apocentric passages.}
\label{fig:stardenspro}
\end{figure*}
 
\begin{figure*}
\includegraphics[width=165mm]{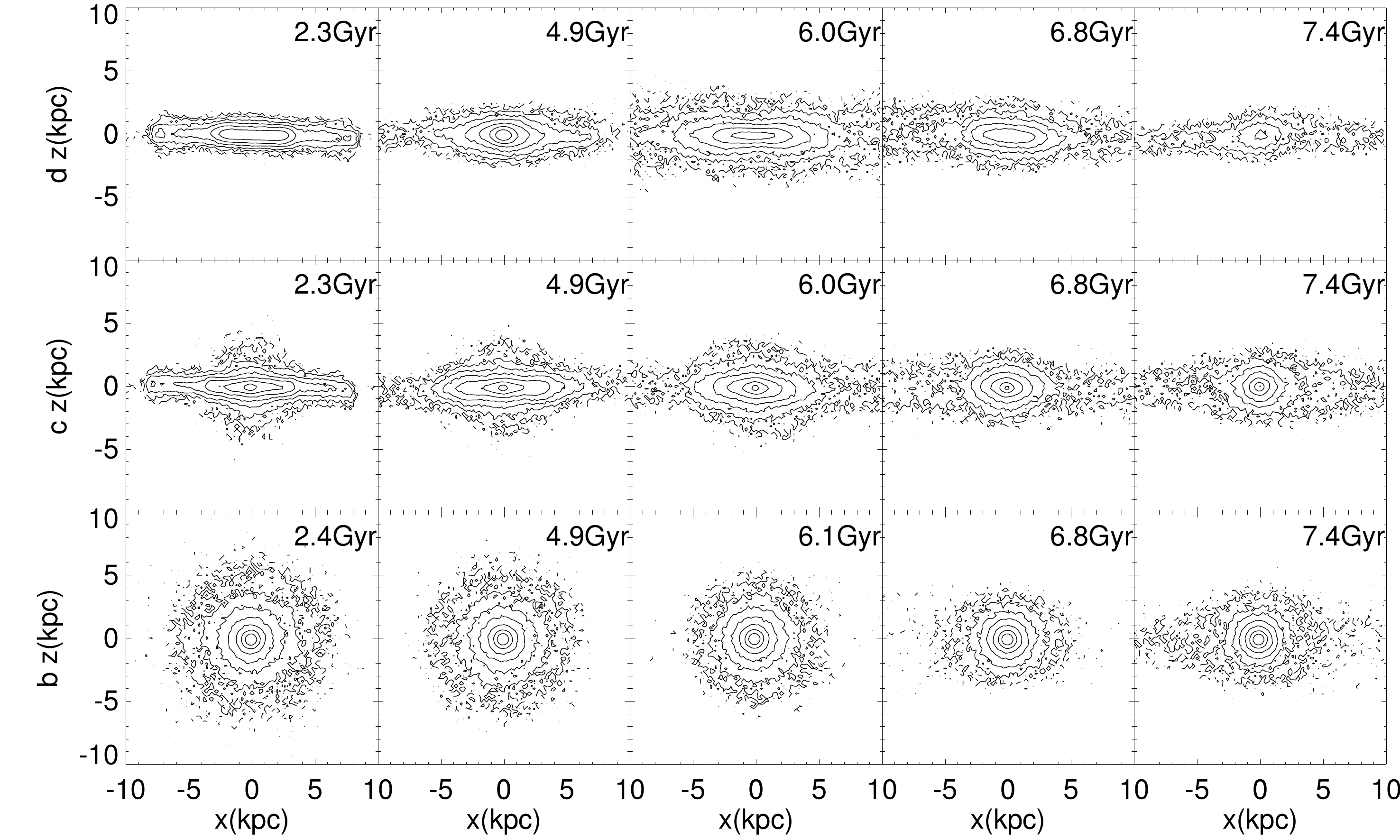}
\caption{Evolution of isodensity contours for the satellite ``d'', ``c'' and
``b''(from top to bottom). The contours are plotted at the same levels.}
\label{fig:morph}
\end{figure*}

As  the  satellite  loses  its  mass due to tidal  stripping,  its  inner
structure   and   morphology  also  change. 
In Fig.\ref{fig:halodenspro} we show the  evolution of dark matter density profile
of the satellite in the C-d, C-c and C-b cases.
During the first 2.5 Gyrs (one pericentric passage) the dark matter
density profile only changes in the external part ($R\gtrsim10$ kpc) with
no effects on the inner part. When the number of pericentric 
passages increases the density profile start to change also in
the inner region ($R\approx 2$ kpc) with a constant dropping
of the central density value. After 6 Gyrs the evolution 
of the dark matter density profile is linked to the evolution
of the stellar component: for composite and bulge satellite, the 
contribution to the overall potential due to the stellar component 
is able to retain more dark matter in the center. On the contrary
for the disc satellites, due to the very quick removal of the stellar component,
that can not help retain matter. In the center, the dark matter distribution
drops by two orders of magnitude in its central density
at $t=7.4$ Gyrs.
This in qualitative agreement with previous results by Kazantzidis 
\etal  (2004, 2011), even if a direct comparison is difficult given
the larger mass ratio they assumed for their (1:100 and 1:1000) satellites
and the different orbital parameters.

Fig.~\ref{fig:stardenspro} shows the  evolution of the stellar 
surface density of the  satellite. 
The  inner surface density of  satellites ``c'' and ``b''  
is practically unchanged during the whole evolution.
The outer part of satellite ``c'' decreases significantly, as a consequence
of the removal of the disc component, and at the end the composite
satellite galaxy transforms into a pure bulge galaxy.
For satellite  ``d'' (left panel), the inner
profile  actually  increases  by  about  20\%  during  the  first  few
pericentric  passages,   but  then decreases  dramatically   at  the  
latest stage of the evolution (7.4 Gyr), with a central density reduced
by more than one order of magnitude. The initial increase in the 
central density in the disc case is due to the formation of a bar-like 
structure after the first passage.

A more clear understanding of the morphological evolution can be seen
in Fig.~\ref{fig:morph}. For the disc case it is evident the formation
of a central concentration (bar-like) after the first passage (4.9 Gyrs),
and the subsequent quick dissolution of the stellar component.
For the composite galaxy the evolution from disc dominated to bulge dominated
is also quite evident, with the disc being completely removed after 
6.8 Gyrs. Finally the bulge case does not show any appreciable change
on the inner stellar distribution, but just a progressively mass loss
in the external part.

\subsection{Dark Matter/Stellar mass evolution}
\begin{figure*}
\includegraphics[width=160mm]{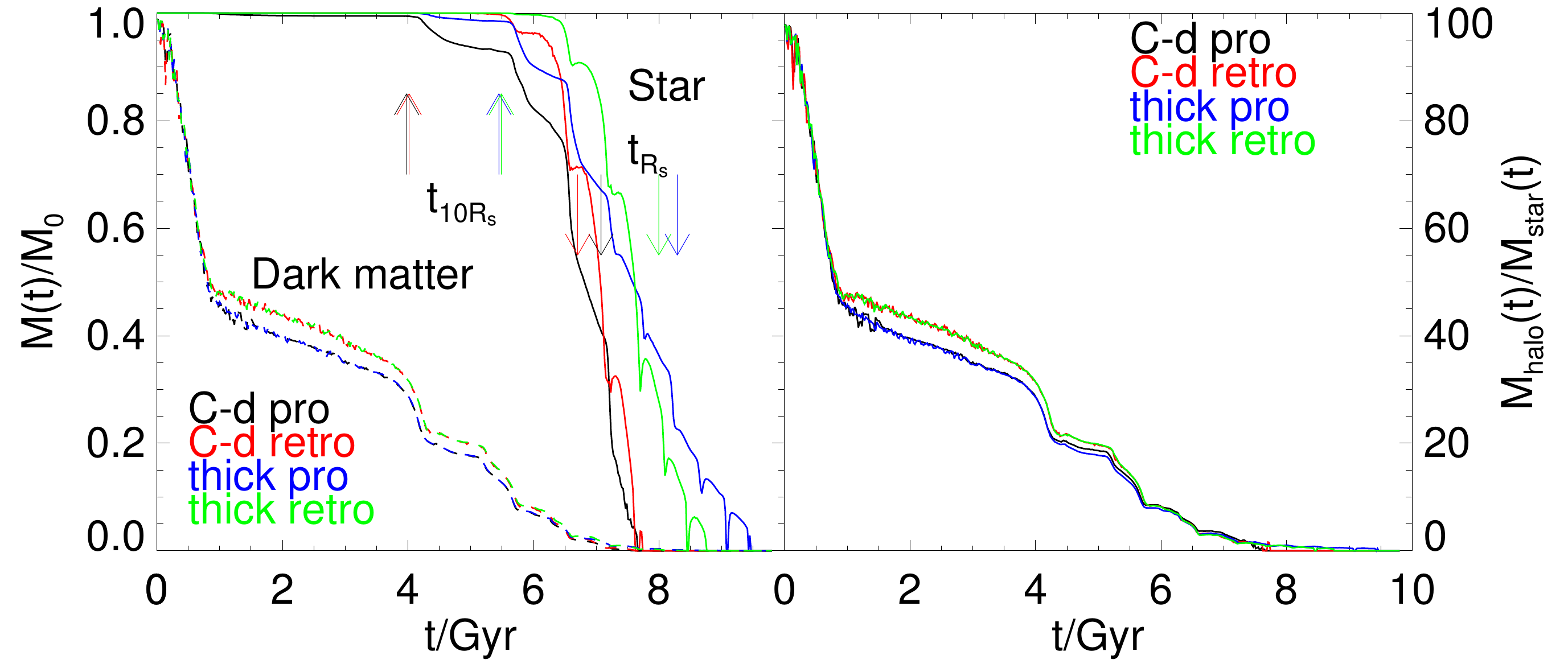}
\caption{The left panel shows the  evolution of bound mass of prograde
  and retrograde  of two  kind of  different morphologies.  The dashed 
  line  is the  dark matter  and solid  line is  the stellar component,  and the
  upward arrows show the time of $t_{10R_s}$ of different mergers. The
  right panel shows the mass ratio between dark matter and stellar mass.}
\label{fig:massratio}
\end{figure*}

Figure \ref{fig:dislogbind3} clearly shows that the dark matter mass
is removed more efficiently than the stellar component, it is then
worth to look at the time evolution of the ration between dark
matter and stellar mass in our simulations.
Recently a new form of stripping (dubbed {\it resonant stripping}, D'Onghia
\etal 2009) has been proposed. In this formalism the disc  angular momentum 
can  couple with the orbit angular momentum, as a consequence
the stellar particles within the  disc can be
stripped  by  the tidal  force  more effeciently than the dark matter,
possibly creating a dark matter dominated satellite.
Since we expect this angular momentum coupling to happen only for prograde 
encounters, we run an additional retrograde C-d simulation.

The evolution of the mass removal and of the dark to stellar ratio
is shown in Figure~\ref{fig:massratio}.
We do not find any evidence for resonant stripping, even if we
do find that prograde mergers are marginally more effective in inducing tidal 
stripping effects as well known in the literature (Cox \etal, 2006).
We also find that the dark matter-to-stellar ratio has a monotonic 
behavior and that the satellite becomes more and more 
stellar dominated as time goes by, due to the removal of the external
dark matter envelop.
It is also worth noticing that our time scales ($t_{10R_s}$ and $t_{R_s}$) 
are still able to characterize the mass loss evolution.

\subsection {Dependence on host morphology}
\begin{figure*}
\includegraphics[width=160mm]{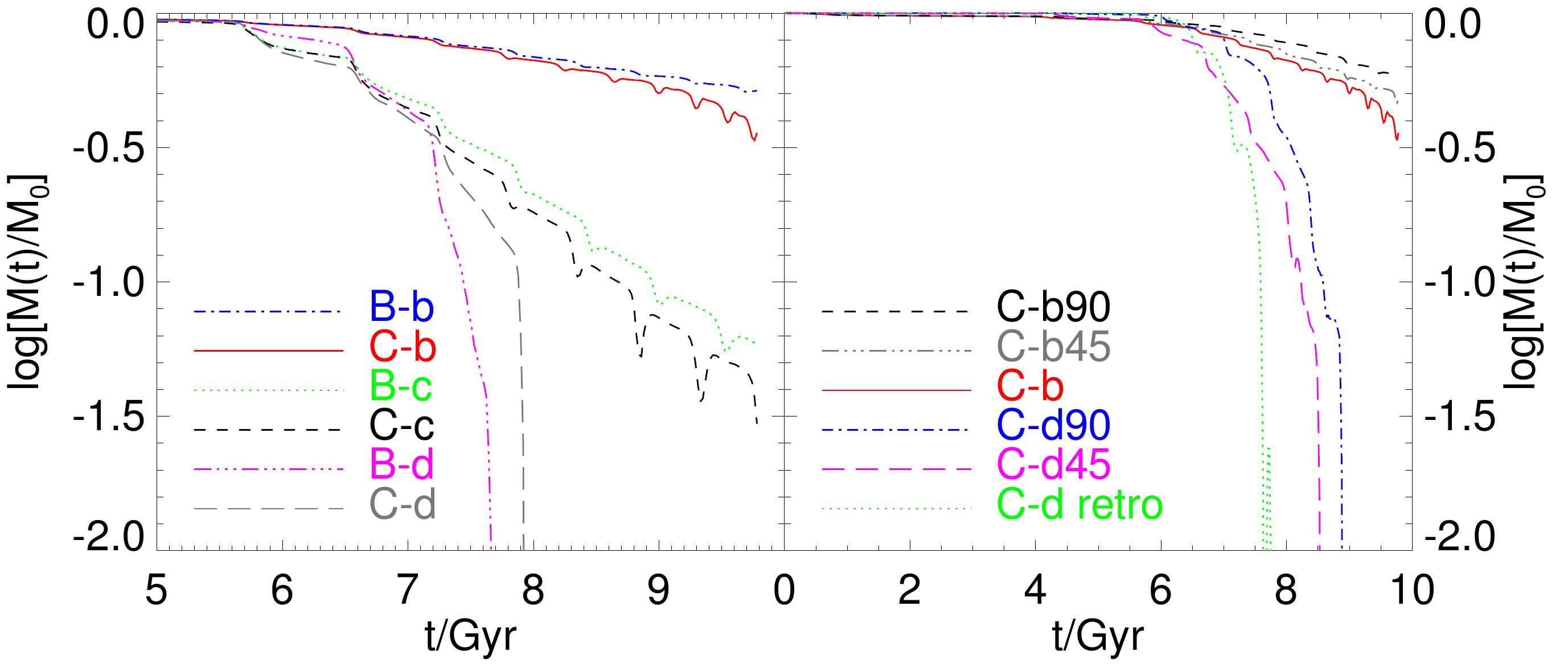}
\caption{The left panel shows the effect of the host morphology on the mass loss of a ``b''
, ``c'' and ``d'' satellite(from top to bottom). The morphology of host
galaxy has a second order of effect on the remnant mass of satellites compared
to the morphology of satellites themselves. The right panel shows the evolution
of the stellar mass for different orbit polar angles $\theta_r$ (see
  Table.\ref{tab:orbit-param}).}
\label{fig:zoomin}
\end{figure*}

In  previous sections, we  have investigated  the dependence  of tidal
stripping on satellite morphology. In this part we examine the effects
of central  host galaxy morphology and the orbit of satellites.  
The effect of the host galaxy morphology on stellar stripping is shown  
in the left  panel of Fig.~\ref{fig:zoomin}. 
For a fixed satellite morphology (composite of bulge) the effect of the central galaxy
is very mild, with the a composite (bulge+disc) morphology being more 
effective in inducing tidal stripping (see for example the difference
between the blue and the black curve, between the cyan and black red or between
the gray and magenta).
This effect is due to the more extended stellar component in the ``C''
case and to its the more asymmetric potential.

\subsection{Dependence on mass ratio and satellite orbit}

\label{sec:sat_orbit}
\begin{figure*}
\includegraphics[width=160mm]{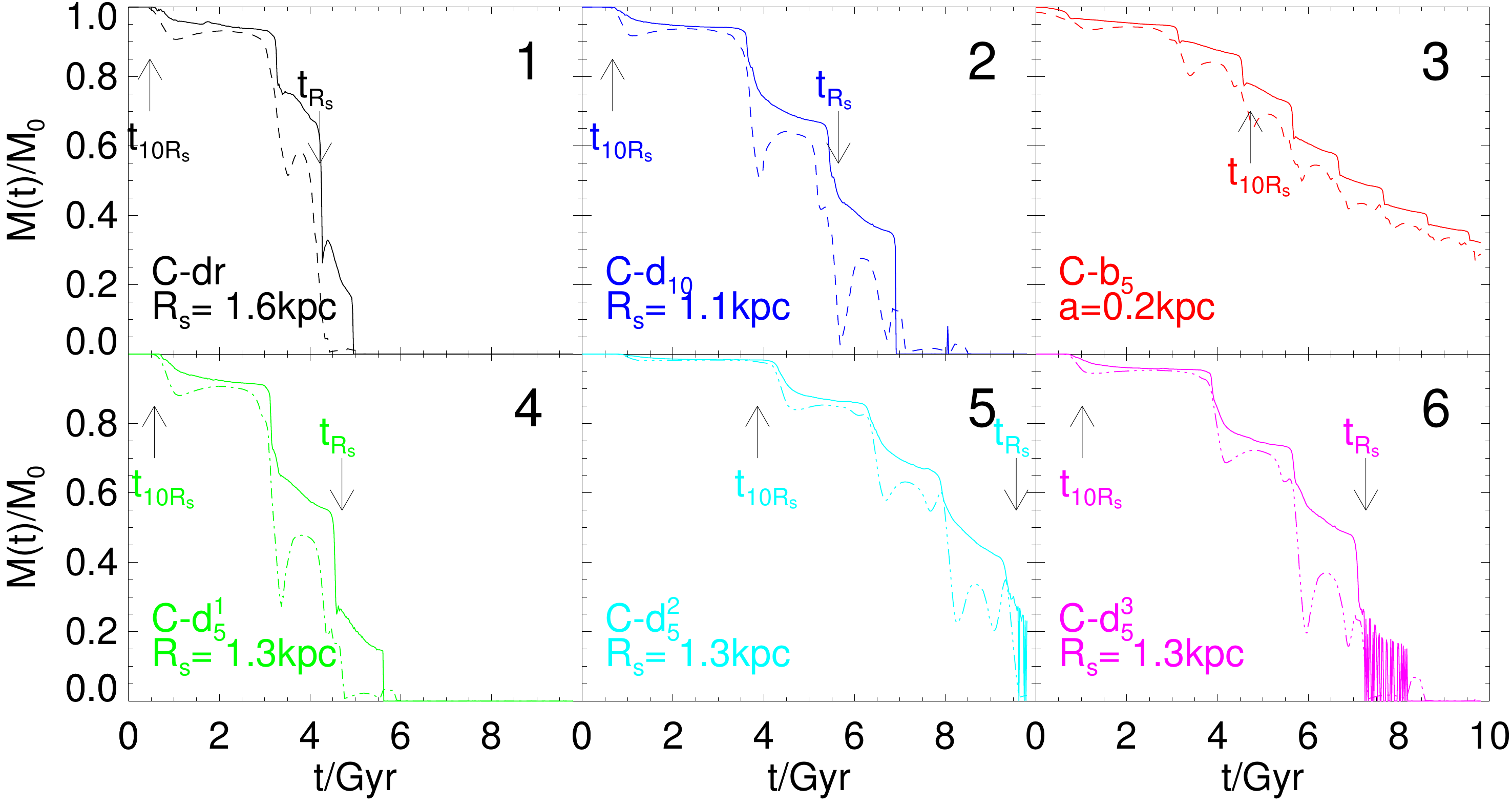}
\caption{Effects of mass ratio and orbital parameters on stellar mass loss. The
dashed lines and dot dashed lines show the tidal mass and the solid lines indicate the stellar
remnants. The upper pannels show the merger C-dr (C-d merger on a more radial orbit),
C-d$_{10}$ (C-d merger with a 10 times lighter satellite) and C-b$_5$ (C-b merger
with a 5 times lighter satellite), and the lower pannels show the merger C-d$_5$ with 3
different orbits (see Table.~\ref{tab:orbit-param}).}
\label{fig:bindtidal6}
\end{figure*}

As found in proviouss works (Villalobos \etal
2012), the mass and the orbit of the satellite make an important role in
its tidal evolution. We use satellites which have 5 times or 10 times less massive stellar component
with respect to the reference satellite to test how the mass ratio affects tidal
stripping and life time. As shown in Fig.~\ref{fig:bindtidal6}, pannel 2, 3, and
4, the different
mass ratios do have effects on the tidal stripping. However, the bound mass
profile still follows the tidal mass profile very well and $t_{10R_s}$ and $t_{R_s}$
still mark the beginning of stellar mass loss, and when the tidal stripping
becomes exponential.

If the central galaxy has a stellar disc, the orbit of the
satellite  may  have   different  inclination, $\theta_{r}$,  with  respect
to this central disc. In the right  panel of Fig.~\ref{fig:zoomin} we plot the
evolution of the stellar mass loss for a satellite in an orbit coplanar 
with the disc ($\theta_{r}=0^{\circ}$, black), 
perpendicular to disc ($\theta_{r}=90^{\circ}$, red),  and  in between    with
$\theta_{r}=45^{\circ}$ (grey). For the disc, we can get the same results (blue,
magenta and cyan). As expected the tidal stripping  is  the strongest (weakest) for  the coplanar 
(perpendicular) orbit. On the other hand this effect
is quite small and the angle between the orbital plane and 
the galactic disc seems to not play a major role in determining the
tidal stripping. This mechanism works for both disk and bulge morphology. 

Finally we test a few cases for satellites on more radial or more circular
orbits, shown in Fig.~\ref{fig:bindtidal6}, pannel1, 4, 5 and 6. Our results are consistent with previous studies (Boylan-Kolchin \etal,
2008) that a more radial (eccentric) merger leads to a faster mass loss. 
On the other hand we do not find any particular trend between the orbit
and the satellite morphology; this implies that our previous results also
apply in the case of different orbits.

\section{Discussion and Conclusions}
\label{sec:conclusions}

In this work, we used N-body simulation to study how the morphology 
of satellite and central galaxy affect the tidal stripping in a series 
of isolate minor mergers.

We kept all parameters of the merger (stellar mass, dark halo mass and orbital
parameters) fixed while we experiment with three different morphologies for the satellite: 
pure disc, pure bulge and bulge+disc and two for the host galaxy: bulge+disc and pure bulge.

In agreement with previous studies we find that the removal of the stellar component
only begins after the satellite dark matter halo has lost a considerable fraction 
of its mass, of the order of 90\%.

We find that for a  fixed stellar mass, the morphology of the satellite has 
a very strong effect in determining its final stellar mass after several 
orbital periods. Depending on its initial stellar distribution a satellite can either
survive to the tidal forces from the central halo retaining a considerable fraction 
of its mass, or being completely disrupted.

We find that there are two characteristic time scales that define the stellar stripping
of a pure disc satellite. The first one sets the beginning of the stellar
stripping (after the removal of the outer dark matter envelope) and happens
when the tidal radius is of the order to ten times the scale radius of the disc.
The second one sets starting of the very quick removal of the disc component
(exponential mass loss) and occurs when the tidal radius is of the order of the
disc scale length.
These two time scales are quite insensitive to the actual value of the 
disc scale lengths and to the merger orbital parameters. They are very 
useful to define the stellar stripping of disc like satellites in models
aiming to describe galaxy formation.

For a composite morphology (disc+bulge) we find that the disc component is quickly 
removed as in the pure disc case and then the remaining bulge is able to survive
for a longer time.

The morphology of the central galaxy has only a second order effect on the tidal stripping,
with a bulge+disc model being slightly more effective in removing dark and stellar mass from
the satellite.

Finally we look at effects of the angle between the orbital plane and the
central galactic disc. As expected satellite moving in the plane of the disc
experiences a 
somehow stronger mass loss than satellite moving perpendicular to the disc. But 
again we find that the effect of the orbit orientation is sub-dominant with respect 
to the satellite morphology.

The effect of satellite morphology on the tidal stripping, usually neglected in the galaxy formation
models, could have important effects on the predictions of the diffuse stellar
light in galaxy groups and clusters, as well as the abundance of disc dominated
satellites. Our results imply that only satellites with a substantial bulge can survive
for a long time, while disc dominated satellites, if still present, should have
been accreted recently. We plan to address in details the effects of
stellar stripping on galaxy formation and evolution in a forthcoming
paper.

\section*{Acknowledgments} 
We kindly thank Volker Springel for providing the code to set up the initial galaxy for our 
simulation run. The numerical simulations used in this work were performed on the THEO cluster 
of the Max-Planck-Institut f\"ur Astronomie at the Rechenzentrum in Garching. 
The authors acknowledge support from the MPG-CAS through the partnership program between 
the MPIA group lead by A. Macci\`o and the PMO group lead by X. Kang.
AVM acknowledges funding from the Deutsche Forschungsgemeinschaft via the SFB 881 program 
"The Milky Way System". Xi Kang is supported by the NSFC (No. 11073055), 
National basic research program of China (2013CB834900), 
the foundation for the author of CAS excellent doctoral dissertation, and the Bairen program of CAS. 
J. Chang acknowledges support of the MPG-CAS student program.

\bibliographystyle{mn.bst}
\bibliography{mn-jour,biblio}

\appendix

\bsp

\label{lastpage}

\end{document}